\documentstyle [12pt]{article}
\advance\hoffset by -.5 in
\advance\voffset by -1 in
\begin{document}

\def\pz{\p_z}
\def\a{\alpha}
\def\b{\beta}
\def\g{\gamma}
\def\d{\delta}
\def\e{\epsilon}
\def\x{\xi}
\def\f{\phi}
\def\j{\psi}
\def\t{\tau}
\def\L{\Lambda}
\def\l{\lambda}
\def\p{\partial}
\def\o{\omega}
\def\O{\Omega}
\def\z{\zeta}
\def\Tau{{\rm T}}
\def\cA{{\cal A}}
\def\tcA{\tilde{\cal A}}
\def\cK{{\cal K}}
\def\tf{\tilde{f}}
\def\tg{\tilde{g}}
\def\th{\tilde{h}}
\def\rarr{\rightarrow}
\def\mp{\mapsto}
\def\hL{\hat{L}}
\def\res{{\rm res}~}
\def\tr{{\rm tr}~}
\def\tres{\tr\res}
\def\cF{{\cal F}}
\def\et{\eta}
\def\cV{{\cal V}}
\def\s{\sum}
\def\cL{{\cal L}}
\def\cH{{\cal H}}
\def\hw{\hat{w}}
\def\bg{{\bf g}}
\def\un{\underline}

\begin{center}
\begin{bf}
\begin{Large}
Additional symmetries of KP, Grassmannian, \\
and the string equation II\\
\end{Large}
\vspace{.5in}
L.A.Dickey\\
\end{bf}
\vspace{.3in}
University of Oklahoma, Norman, OK 73019 \\
e-mail: ldickey@nsfuvax.math.uoknor.edu \\
\vspace{.3in}
October, 1992
\end{center}
\vspace{.5in}
{\bf 1. Introduction.} As in the first part of this paper (see [1]) we are
interested in the connection between additional symmetries and the string
equation. Now we extend this theory to the multi-component (i.e. matrix) KP or
KdV (mcKdV) hierarchies, and to modified KdV (mKdV).

In addition to having pure mathematical interest, this study is also necessary
for some physical applications. It is known that some matrix models, such
as e.g. unitary matrix models, lead to the string equation for mKdV (see [10]
and [2]). An important feature of a string equation is
its close relation to some hierarchy of integrable equations. This is
well-known
in the case of the string equation of KdV hierarchies. The relation between
them consists in the fact that the string equation is invariant under the flows
generated by the hierarchy equations. As the experience with KdV shows, the
origin of this invariance is the following: there are some additional
symmetries of the hierarchy which do not belong to the hierarchy although they
commute with it. The set of fixed points of an additional symmetry is
invariant under the hierarchy, and the string equation is nothing but the
condition of this stationarity.

Thus, we would study additional symmetries acting on mKdV if not for a little
obstacle: to all appearances, the additional symmetries do not act on
mKdV at all! How can this happen if it is known that they act on KdV
hierarchies, and the latter are connected with mKdV by the Miura
transformation? The problem is that there are, in fact, $n$ different KdV
operators corresponding to the same mKdV operator, and the results of action
of the additional symmetry on them are not compatible (this is only our
conjecture but a very plausible one).

Fortunately, the mKdV equation can be embedded into the mcKdV hierarchy (we are
not sure if this fact has been somewhere mentioned before). In this, wider,
phase space the additional symmetries do not act worse than they did for the
ordinary, scalar KdV. It is true that the flows of additional symmetries do not
respect the submanifold representing mKdV. But if a point on the submanifold is
stationary under an additional symmetry, it stays on this submanifold. The mKdV
string equation induces the corresponding KdV string equation.

Irrespectively of their application to mKdV, the mcKdV can play a role in some
matrix models. This is why we start from these hierarchies. In contrast to [10]
or [2] we consider hierarchies of an arbitrary order $n$ (instead of $n=2$):
consideration of a special case often obscures general facts and formulas.

Here, as well as in the first part, we are prompted by a desire to show that
the notion of an additional symmetry which, only a few years ago, seemed to be
some peripheral part of the theory of integrable systems is actually the base
of these new developments (string equation, Virasoro constraints or, more
general, W-constraints). We use the most general and convenient form of
additional symmetries given in [4] (while the first work introducing them was
[5]). Acquaintance with the general background of this theory can be made
e.g. with the help of [6].

As in the first part, we also study the action of additional symmetries on the
Grassmannian.\\

{\bf Part 1. Multi-component KP and KdV (mcKdV).}\\

{\bf 2. Definition.} Let
$$ L=A\partial +u_0+u_1\partial^{-1}+\cdots$$
where $u_i$ are $n\times n$ matrices, $A={\rm diag}(a_1,...,a_n),\:\:a_i$ are
distinct non-zero constants. Diagonal elements of $u_0$ are assumed to be zero.
Let $R_{\alpha}=\sum_{j=0}^{\infty}R_{j\alpha}\partial^{-j}$, $\alpha=1,...,n,$
where $R_{0\alpha}=E_{\alpha}$, $E_{\alpha}$ is a matrix having only one
non-zero element on the $(\a,\a)$ place which is equal to 1; $R_{\a}$ is
supposed to satisfy $$ [L,R_{\alpha}]=0.$$
It can be shown that such matrices exist and their elements are differential
polynomials in elements of $u_i$ being $$R_{\alpha}R_{\beta}=\delta_{\alpha
\beta}R_{\alpha},\hspace{.1in}\sum_{\a=1}^n R_{\alpha}=I$$ (i.e. this is a
spectral decomposition of the unity). The mcKP hierarchy (multicomponent KP) is
$$\p_{k\alpha}L=[(L^kR_{\alpha})_+,L],~~\p_{k\a}=\p/\p t_{k\a},~~k=0,2,...;~\a=
1,...,n \eqno{(2.1)}$$ and $t_{k\a}$ are the ``time variables" of the
hierarchy. The equations imply $\p_{k\a}R_{\b}=[B_{k\a},R_{\b}]$. Notice that
$\p_{0\a}L=[E_\a,L]$.

It can be shown that the equations commute. The variables $x$ and $t_{k\alpha}
$ are not independent: $$\partial=\sum_{\alpha}a_{\alpha}^{-1}\partial_{1\alpha
}$$ (Greek indices always run from 1 to $n$).\\

{\bf Remark 1.} {\sl Two viewpoints on the last formula are possible: i) not to
introduce explicitly $x$ and to consider $\p$ as $\sum_\a a_\a^{-1}\p_{1\a}$ by
definition, ii) to consider a separate variable $x$, however, $x$ and $t_{1\a}$
are involved in formulas not independently but only in combinations $t_{1\a}+
a_\a^{-1}x$. Here we accept the first point, in Sects. 7 and 8 the second one.}
Let $$L=\hw A\p\hw^{-1},~~{\rm where}~~\hw=\hw(A\p)=\sum_0^{\infty}w
_i(A\p)^{-i},~w_0=I;$$ Then $R_{\alpha}=\hw E_{\alpha}\hw^{-1}.$ Put
$$w=\hw(A\p)\exp\xi(t,z)=\hw(z)\exp\xi(t,z);~~{\rm where}~~\xi(t,z)=\sum_{k=
0}^{\infty}\sum_{\alpha=1}^nz^kE_{\a}t_{k\alpha}.$$
This is the {\em Baker function}; it satisfies the equations
$$Lw=zw,\hspace{.1in}{\rm
and}\hspace{.1in}\partial_{k\alpha}w=(L^kR_{\alpha})_+
w. $$ The latter equation is equivalent to
$$ \partial_{k\alpha}\hw=-(L^kR_{\alpha})_-\hw.\eqno{(2.2)}$$
{\bf  Remark 2}. {\sl Series $\hw$ are defined up to a multiplication on the
right by series $\sum_0^\infty a_i\p^{-i}$ with constant diagonal matrices
$a_i$. Instead of being constant they can satisfy a weaker condition $\p a_i=0
$, then the last equation must be written in a more general form}:
$$ \partial_{k\alpha}\hw=-(L^kR_{\alpha})_-\hw+\hw b,$$
{\sl where $b=\sum_1^\infty b_i\p^{-i}$ is a diagonal series in $\p^{-1}$ with
coefficients not depending on $x$}.\\

{\bf Remark 3.} {\sl We have $\p_{0\a}\hw=-(R_\a-E_\a)\hw=-\hw E_a+E_\a\hw=
[E_\a,\hw].$ Symmetries related to ``zero" time variables $t_{0\a}$ are
similarity transformations with constant matrices.}\\

{\bf 3. Additional symmetries.} We introduce them in a way similar to that in
the scalar case (see, e.g. [6], sect. 7.8.1.).

The operator $\p_{k\a}-(L^kR_\a)_+$ can be
represented in a ``dressing" form: $$\p_{k\a}-(L^kR_\a)_+=\hw\p_{k\a}\hw^{-1}
+(\p_{k\a}\hw)\hw^{-1}-(L^kR_\a)_+$$ $$=\hw\p_{k\a}\hw^{-1}-L^kR_\a=\hw(
\p_{k\a}-(A\p)^kE_\a)\hw^{-1} \eqno{(3.1)}$$ (using (2.2)). Thus, the
hierarchy equation (2.1) is a dressing of an obvious equality \break
$[\p_{k\a}-(A\p)^kE_\a,A\p]=0$.

There is another operator commuting with $\p_{k\a}-(A\p)^kE_\a$. This is
$$\Gamma=\sum_{l=1}^\infty\sum_{\b=1}^nt_{l\b}l(A\p)^{l-1}E_\b.$$ Put
$M=\hw\Gamma\hw^{-1}$. Dressing the equality $$[\p_{k\a}-(A\p)^kE_\a,\Gamma]=
0$$ we obtain $$\p_{k\a}M=[(L^kR_\a)_+,M].$$ Together with (2.1) this
implies more general equality $$\p_{k\a}(M^kL^m)=[(L^kR_\a)_+,M^kL^m].
\eqno{(3.2)}$$ In order to define a one-parameter group of additional symmetry,
take some variable $t_{l,m,\a}^*$ (the asterisk superscript will distinguish
variables of additional symmetries) and consider an equation $$\p_{l,m,\a}^*
\hw=-(M^lL^mR_\a)_-\hw,~~\p_{l,m,\a}^*=\p/\p t_{l,m,\a}^* \eqno{(3.3)}$$
which is the definition of a one-parameter group of transformation of $\hw$.
Then $$\p_{l,m,\a}^*L=-[(M^lL^mR_\a)_-,L],~~\p_{l,m,\a}^*R_\a=-[(M^lL^mR_\a)_-,
R_\a].\eqno{(3.4)}$$ This implies
$$\p_{l,m,\a}^*(L^kR_\b)_-=-[(M^lL^mR_\a)_-,(L^kR_\b)_-]_--[(M^lL^mR_\a)_-,
(L^kR_\b)_+]_-$$ $$=-[(M^lL^mR_\a)_-,(L^kR_\b)_-]-[M^lL^mR_\a,\p_{k\b}]_-=-[(M^
lL^mR_\a)_-,(L^kR_\b)_-+\p_{k\b}].$$
{\bf Proposition 1.} {\sl $\p_{k\b}$ and $\p_{l,m,\a}^*$ commute}.
$$[\p_{l,m,\a}^*,\p_{k,\b}]L=-\p_{l,m,\a}^*[(L^kR_\b)_-,L]+\p_{k,\b}[(M^lL^mR_
\a)_-,L]$$ $$=[\p_{k,\b}(M^lL^mR_\a)_--\p_{l,m,\a}^*(L^kR_\b)_-,L]+[(L^kR_\b)_-
,[(M^lL^mR_\a)_-,L]]$$ $$-[(M^lL^mR_\a)_-,[(L^kR_\b)_-,L]]=[\p_{k,\b}(M^lL^mR_
\a)_--\p_{l,m,\a}^*(L^kR_\b)_-,L]$$ $$+[[(L^kR_\b)_-,(M^lL^mR_\a)_-],L]$$ $$=[
\p_{k,\b}(M^lL^mR_\a)_--\p_{l,m,\a}^*(L^kR_\b)_-+[(L^kR_\b)_-,(M^lL^mR_\a)_-],L
]$$ $$=[[\p_{k,\b}+(L^kR_\b)_-,(M^lL^mR_\a)_-]-\p_{l,m,\a}^*(L^kR_\b)_-,L]$$
$$=[[\p_{k,\b}+(L^kR_\b)_-,(M^lL^mR_\a)_-]+[(M^lL^mR_\a)_-,\p_{k,\b}+(L^kR_\b)_
-],L]=0.$$ If they commute on $L$ then they commute on the whole differential
algebra generated by coefficients of $L$.\\

{\bf Proposition 2.} {\sl The Lie algebra of operators $\p_{l,m,\a}^*$ is
isomorphic to the Lie algebra generated by $-z^l\pz^mE_\a$ where $\pz=\p/\p
z$ (or $-z^m\pz^lE_\a$, which is the same).}\\
This is what they call $W_\infty$-algebra.

{\em Proof.} First we find the following commutation relations:
$$ [L,M]=\hw[A\p,\sum_{l,\b}t_{l\b}(A\p)^{l-1}E_\b]\hw^{-1}=\hw{1
\over n}\sum_\b a_\b\cdot a_\b^{-1}E_\b\hw^{-1}=1,\eqno{(3.5)}$$ $$ [R_\a,M]=
\hw[E_\a,\sum_{l,\b}t_{l\b}(A\p)^{l-1}E_\b]\hw^{-1}=0.\eqno{(3.6)}$$ Then
we prove\\

{\bf Lemma.} {\sl The correspondence $\p_{l,m,\a}^*\mapsto M^lL^mR_\a$ is an
anti-isomorphism of Lie algebras.}\\

(One may not identify $\p_{0,m,\a}^*$ and $\p_{m,\a}$ in this lemma: they act
on $M$ in different ways, $\p_{0,m,\a}^*M=-[(L^mR_\a)_-,M]$ while $\p_{m,\a}M
=[(L^mR_\a)_+,M]$; the action of these operators on $L$ coincides. The reason
of
this distinction is the explicit dependence $M$ of $t_{k\a}$).

{\em Proof of the lemma}. For the simplicity of notations, let $M^lL^mR_\a=
a_{lm\a}$. We have (a circle around a subscript ``minus" ,$\ominus$, means that
this subscript can be omitted)
 $$[\p_{l,m,\a}^*,\p_{k,n,\b}^*]L=-\p_{l,m,\a}^*[(a_{kn\b})_-,L]-
((l,m,\a)\Leftrightarrow (k,n,\b))$$ $$=[[(a_{lm\a})_-,a_{kn\b}]_-,L]+
[(a_{kn\b})_-,[(a_{lm\a})_-,L]]-((l,m,\a)\Leftrightarrow (k,n,\b))$$ $$=
[[(a_{lm\a})_-,(a_{kn\b})_-]_\ominus,L]+[[(a_{lm\a})_\ominus,(a_{kn\b})_+]_-,L]
$$ $$+[[(a_{kn\b})_-,(a_{lm\a})_-],L]+[(a_{lm\a})_-,[(a_{kn\b})_-,L]]$$ $$
-[[(a_{kn\b})_-,a_{lm\a}]_-,L]-[(a_{lm\a})_-,[(a_{kn\b})_-,L]]$$ $$
=[[a_{lm\a},a_{kn\b}]_-,L]=-[[a_{kn\b},a_{lm\a}]_-,L].~~\Box$$ The rest follows
from the fact that the
commutation relation for $M$ and $L$ is the same as for $z$ and $\pz$, that
$\{R_\a\}$ commute with both $L$ and $M$, and that $\{R_\a\}$ is a spectral
decomposition of the unity. The proposition is proven.\\

{\bf Remark.} {\sl The isomorphism stated in the proposition 2 is not simply
an abstract isomorphism: operators $-z^m\pz^lE_\a$ are exactly how $\p_{l,m,\a
}^*$ act on the Grassmannian (see sect. 8).}\\

{\bf 4. Reductions.} The hierarchy and additional symmetries admit reductions
to submanifolds of operators $L$ such that $L^h$ is a purely differential
operator, i.e. $L^h=L_+^h$, operators $L^k$ being understood as
$(L_+^h)^{k/h}$.
This means that the right hand sides of the equations giving the hierarchy and
additional symmetries can be written in terms of $(L_+^h)^{k/h}$ modulo
$L_-^h$.
This implies that, if initially $L_-^h=0$, it will not arise in the process of
motion along the trajectories of the hierarchy or of the additional symmetries.

We shall call this reduction the $h$th mcKdV reduction. For this reduction we
have $$\sum_\a \p_{(lh)\a}L=-[L^{lh},L]=0,~~l=1,2,... \eqno{(4.2)}$$ One of the
most important reduction (considered in [2]) is for $h=1$:
$$ L=A\p+u_0. \eqno{(4.1)}$$ In the scalar case this reduction is trivial.\\

There is also a reduction of a different kind. We can assume that operators
$L$ are formally skew-symmetric, $L^*=-L$ (where $(f\p^k)^*=(-\p)^kf$) which
corresponds to unitary $\hw$. Then $R_\a$ are self-adjoint.
We also can consider only real $L$.

For example, let the matrix dimension $n$ be 2. Consider both reductions, $L$
being real skew-symmetric and $h=1$. Then $L=A\p+u_0$ where $A$ can be chosen
as
$diag(1,-1)$ and $$L=\left(\begin{array}{rr}\p&-v\\v&-\p\end{array}\right).$$
Changing the frame we obtain
$$L=\frac 12\left(\begin{array}{cc}1&1\\1&-1\end{array}\right)\left(
\begin{array}{cc}\p&-v\\v&-\p\end{array}\right)\left(\begin{array}{cc}1&1\\
1&-1\end{array}\right)=
\left(\begin{array}{cc}0&\p+v\\ \p-v&0\end{array}\right).$$ This is the
operator discussed in [2].\\

{\bf 5. String equation.} Let $l\geq 1$. The equation (3.4) implies
$$\p_{l,m,\a}^*L^h=-[(M^lL^mR_\a)_-,L^h]=-[M^lL^mR_\a,L^h]+[(M^lL^mR_\a)_+,L^h]
.$$ We shall assume the system being reduced to purely differential operators
$L^h$. The first term of the r.h.s. of the above formula is $$\sum_{i=1}^l{h
\choose i}{l\choose i}i!M^{l-i}L^{h-i+m}R_\a.$$ If $l=1$
and $m=-h+1$ then the sum reduces to one term which is a constant, $h$. Summing
over $\a$, we have $$\p_{1,-h+1}^*L^h=h+[(ML^{-h+1})_+,L^h].$$ where
$\p_{l,m}^*=\sum_\a\p_{l,m,\a}^*$. Evidently, $\p_{l,m}^*L=-[(M^lL^m)_-,L]$.

The {\em string equation} is $\p_{1,-h+1}^*L^h=0$, or, in an equivalent form,
$$ [L^h,(ML^{-h+1})_+,L^h]=h.$$ In the case of the reduction $h=1$ it is simply
$$[L,M_+]=1$$ (which is an equation, in contrast to the identity $[L,M]=1$).\\

{\bf 6. Grassmannian.}  {\em Definition of the Grassmannian}. Let $H$ be
$L^2(S,
{\bf C}^n)$, a space of series $v(z)=\sum_{-\infty}^{\infty}v_kz^k$
where $v_k\in{\bf C}^n,\:\: |z|=1$, let $H_+$ and $H_-$ be spaces of
truncated series: $H_+=\{\sum_0^{\infty}\},\:\:H_-=\{\sum_{-\infty}^{-1}\}$;
$H=H_+\oplus H_-$. Let $p_{\pm}$ be natural projections of $H$ onto these
subspaces. We shall think of vectors as vector-rows.

The Grassmannian, Gr, is the set of all subspaces $V\subset H$ enjoying the
properties: $p_+|_V$ is a one-to-one correspondence (i.e. we restrict the
definition to a generic case), and $p_-|V$ is an operator of a trace class. We
say that a matrix function belongs to $V$ if
all its rows do. Let $g(t,z)=\exp\xi=\exp\sum_{k=0}^{\infty}\sum_{\a=1}^nz^kE_
{\a}t_{k\a}$. Let ${\bf g}^{-1}$ be a transformation of the space $H$:
$$ {\bf g}^{-1}:H\rightarrow H,\:v\mapsto vg^{-1}.$$
If $V\in$ Gr then for almost all $t$ the subspace $Vg^{-1}\in$ Gr.

The {\em Baker function} corresponding to an element $V\in$Gr, denoted as
$w_V(t,z)$, is determined by conditions
i) for every set of variables $t=(t_{k\a})$ $w_V$ belongs to $V$ as a function
of $z$, ii) $p_+(w_Vg^{-1})=1$. Together the conditions mean that $w_V(t,z)$ is
the unique element of $V$ of the form
$$ w_V(t,z)=(1+\sum_1^{\infty}w_iz^{-i})g(t,z)\equiv\hw_V(t,z)g(t,z).$$

{\bf Proposition 1.} {\sl The Baker function $w_V(t,z)$ is a Baker function of
the mcKP hierarchy defined in a formal sense in the section 2.}\\

The proof can be found in [3].   \\

{\bf Proposition 2.} If $z^hV\subset V$ for an element of the Grassmannian then
$L^h$ is a purely differential operator.\\

{\em Proof}. We have $$\sum_\a(L^hR_\a)_+w_V-z^hw_V=\sum_\a\p_{h\a}w_V-z^hw_V=
\sum_\a\p_{h\a}\hat{w_V}\cdot g(t,z).$$ Now, $\p_{h\a}w_V$ belong to $V$ as
derivatives with respect to parameters, so does $z^hw_V$ by the assumption,
which implies that $\sum_\a\p_{h\a}\hat{w_V}\cdot g(t,z)\in V$. By the above
mentioned uniqueness this gives $\sum_\a\p_{h\a}\hat{w_V}=0$ and
$ \sum_\a(L^hR_\a)_+w_V-z^hw_V=0$ i.e. $(L^h)_+w_V-z^hw_V=0$. On the other
hand, $L^hw_V-z^hw_V=0$ which yields $(L^h)_-w_V=0$ and $L_-^h=0$, as required.

In particular, $L$ is a differential operator, $L=A\p+u_0$ if the corresponding
element of the Grassmannian enjoys the property $zV\subset V$.\\

{\bf 7. $z$-operators and the Grassmannian.} If a reader is not interested in
the problem how the additional symmetries act on the Grassmannian, he can skip
this and the next sections and go directly to the section 9 dealing with action
of additional symmetries on $\tau$-functions and with Virasoro constraints; the
more so, as for the special case of $\p_{1,m,\a}^*$ with $m\leq 0$, responsible
for the string equation, the main result of these two sections, the proposition
of Sect. 8, will be proven independently and much easier in Sect. 9.\\

In this and the next sections we use separate variables $x$ and $t_{1\a}$ (see
Remark 1 in Sect. 2).

In the same way as in the scalar case one can introduce ``$z$-operators",
following Mulase [6]: $$ G=G(\pz,z)=\sum_{j\geq 0,i\leq i_0}a_{ij}z^i\pz^j,~~\p
z=\p/\p z$$ where the coefficients $a_{ij}$ are matrices now. Operators $$G=1+
\sum_{j\geq0,i<0}a_{ij}z^i\pz^j$$ are called monic.

The ring of $z$-operators is isomorphic to the ring of $\Psi$DO under the
mapping $$z\mp -A\p,~~ \pz\mp A^{-1}x$$ since $[\pz,z]=[A^{-1}x,-A\p]=1$.

In the vector case there are the same two Mulase's theorems as in the scalar
case:       \\

{\bf Proposition 1.} {\sl If $V\in$Gr is an element of a Grassmannian then
there is a monic $z$-operator $G$ such that $V^T=GH_+^T$. (The superscript $T$
is here for transpose; elements of $V$ and $H_+$ are vector-rows and those of
$V^T$ and $H_+^T$ are vector-columns).}   \\

{\bf Proposition 2.} {\sl If a $z$-operator preserves $H_+$ then it must
involve only non-negative powers of $z$.} \\

The $z$-operators enable one to find direct relations between elements of the
Grassmannian and the dressing operator $\hw$.

Let $V$ be an element of the Grassmannian. Then, generically, $V\exp(-\xi
(t,z))$ is also an element of the Grassmannian. The proposition 1 implies that
there is a monic operator $\Psi(t,A\pz,-A^{-1}z)$ such that
$$e^{-\xi(t,z)}V^T=\Psi H_+^T.\eqno{(7.1)}$$ For every $v\in V$ there is some
$h_+\in H_+$ such that $\exp(-\xi)v^T=\Psi h_+^T$ i.e. $h_+^T=\Psi^{-1}
\exp(-\xi)v^T$. Differentiating this with respect to $t_{k\a}$ we have $$\p_{k
\a}h_+^T=-\Psi^{-1}(\p_{k\a}\Psi)\Psi^{-1}\exp(-\xi)v^T-\Psi^{-1}z^
kE_\a\exp(-\xi)v^T$$ $$=-\Psi^{-1}(\p_{k\a}\Psi)h_+^T-\Psi^{-1}z^kE_\a
\Psi h_+^T.$$
 Put $L^*=\Psi^{-1}z\Psi$ and $R_\a^*=\Psi^{-1}E_\a\Psi$.
Then the obtained equation can be rewritten as $$[-\Psi^{-1}(\p_{k\a}\Psi)
-((L^*)^kR_\a^*)^T]h_+^T=\p_{k\a}h_+^T.$$ The
right-hand side belongs to $H_+$, then so does the left-hand side. The element
$h_+\in H_+$ is an arbitrary element of $H_+$; applying the proposition 2 we
see that $-\Psi^{-1}(\p_{k\a}\Psi)-((L^*)^kR_\a^*))_-=0$ and
$\p_{k\a}\Psi=-\Psi
((L^*)^kR_\a^*)_-$. Passing to $\Psi$DO we get the same equality and $L^*=
\Psi^{-1}(-A\p)\Psi,~R_\a^*=\Psi^{-1}E_\a\Psi$.

Let us pass now to conjugate operators which means that all matrices should be
replaced by transposes, $\p$ by $-\p$, and all factors written in the
inverse order. Conjugation is denoted by a star, $*$. Let
$$\hw(t,x,\p)=\Psi^*(t,x,\p),~L=(L^*)^*,~R_\a=(R_\a^*)^*.$$
Then we arrive at the equations
$$ \p_{k\a}\hw=-(L^kR_\a)_-\hw,~L=\hw A\p\hw^{-1},~R_\a=\hw E_\a\hw^{-1}.
$$ This is nothing but the equation of the hierarchy. Thus, we constructed a
dressing operator corresponding to $V\in$Gr.

Notice that in our construction variables $t_{k\a}$ and $x$ where independent;
the hierarchy equations show that $\hw$ depends on the variables $x$ and $t_{1
\a}$ only in combinations $t_{1\a}+a_\a^{-1}x$.    \\

{\bf 8. Additional symmetries and the Grassmannian.} We are going to obtain
the action of additional symmetries (3.3) on elements of the Grassmannian.

All operators involved in (3.3) depend on $x$ in combinations $t_{1\a}+a_\a^
{-1}x$ (see Sect. 2, Remark 1). In particular, $M=\hw\Gamma\hw^{-1}$ where
$$\Gamma=\sum_{l=1}^\infty\sum_{\b=1}^nt_{l\b}l(A\p)^{l-1}E_\b+\sum_{\b=1}^n
a_\b^{-1}xE_\b=\sum_{l=1}^\infty\sum_{\b=1}^nt_{l\b}l(A\p)^{l-1}E_\b+A^{-1}x.$$
Let us take the conjugated equation: $$\p_{l,m,\a}^*\Psi=-\Psi((L^*)^m(M^*)^lR_
\a^*)_-$$ where, as in the previous section, $\Psi=\hw^*$.

Now, pass from the $\Psi$DO to the $z$-operators. A relation $\exp(-\xi)v^T=
\Psi h_+^T$ connects an arbitrary element $v$ of $V\in$Gr with an element
$h_+\in H_+$. Act with $\p_{l,m,\a}^*$ on both the sides of this equality:
$$\p_{l,m,\a}^*\exp(-\xi)v^T=-\Psi((L^*)^m(M^*)^lR_\a^*)_-h_+^T+\Psi\p_{l,m,\a
}^*h_+^T$$ $$=-\Psi((L^*)^m(M^*)^lR_\a^*)h_+^T+\Psi g_+^T$$ where $g_+^T=((
L^*)^m(M^*)^lR_\a^*)_+h_+^T+\p_{l,m,\a}^*h_+^T$ is an element of $H_+$.
Now, $$\p_{l,m,\a}^*\exp(-\xi)v^T=-z^m
(\sum_{l=1}^\infty\sum_{\b=1}^nt_{l\b}l(A\p)^{l-1}E_\b+\pz)^lE_\a\Psi
h_+^T+\Psi
g_+^T$$
$$=-z^m(\sum_{l=1}^\infty\sum_{\b=1}^nt_{l\b}l(A\p)^{l-1}E_\b+\pz)^lE_\a
\exp(-\xi)v^T+\exp(-\xi)v_1^T$$ where $v_1$ is an element of $V$. Now, notice
that
$(\sum_{l=1}^\infty\sum_{\b=1}^nt_{l\b}l(A\p)^{l-1}E_\b+\pz)\exp(-\xi)=0,$
hence
$$(\sum_{l=1}^\infty\sum_{\b=1}^nt_{l\b}l(A\p)^{l-1}E_\b+\pz)^l(\exp(-\xi)v^T)=
\exp(-\xi)\pz^lv^T.$$ Using this formula and canceling $\exp(-\xi)$ in the
previous equation we obtain
$$\p_{l,m,\a}^*v^T=-z^m\pz^lE_\a v^T+v_1^T.$$ Since the action of an
infinitesimal operator on an element of the Grassmannian is defined as a
mapping $V\rarr H/V$ (see e.g. [1]), the second term does not play any role,
and the operator $\p_{l,m,\a}^*$ acts on elements of the Grassmannian as the
operator $-z^m\pz^lE_\a$. Thus, we have proven:\\

{\bf Proposition.} {\sl Additional symmetries $\p_{l,m,\a}^*$ act on
transposed elements of the Grassmannian (i.e. on vector-columns) as operators
$-z^m\pz^lE_\a$. (Their action on vector-rows differs: multiplication on $E_\a$
is on the right).}  \\

If an operator ${\bf E}_\a$ acting on elements $v\in V$ as ${\bf E}_\a v=vE_\a$
is introduced then the above proposition can be formulated as: $\p_{l,m,\a}$
act on the elements of the Crassmannian as operators $-z^m\p^l{\bf E}_\a$.\\

{\bf 9. Action of additional symmetries on the Baker and the $\tau$-functions.}
Like the Baker
functions of the usual KP, Baker functions of mcKP can be expressed in terms of
$\tau$-functions (see, e.g., [3]). The significance of this fact is that
infinitely many variables, coefficients of the Baker function, are replaced by
only one function, in the case of KP, and $n^2$ functions for mcKP.

In the latter case a $\tau$-function is a matrix $\Tau=\{\tau_{\a\b}\}$ with
equal diagonal elements $\tau_{\a\a}=\tau$, and the relation between the Baker
and the $\tau$-functions is the following:
$$\hw_{\a\a}(t,z)={\t(...,t_{s\g}-\delta_{\g\b}\cdot 1/sz^s,...)\over\t(t)},
\eqno{(9.1)}$$ and $$\hw_{\a\b}(t,z)=z^{-1}{\t_{\a\b}(...,t_{s\g}-\delta_{\g\b}
\cdot 1/sz^s,...)\over\t(t)},~\a\neq\b.\eqno{(9.2)}$$ ({\em Attention: only
those time variables $t_{s\g}$ are shifted whose index $\gamma$ coincides with
the number of the column,} $\g=\b$).

We start with the action of additional symmetries on the Baker functions and
then transfer it to the $\tau$-functions, using (9.1) and (9.2). For simplicity
we are doing this for the most important case of symmetries $\p_{1,m,\a}^*$,
related to the string equation.

For the dressing operator $\hw=\hw(A\p)$ we have $$\p_{1,m,\a}^*\hw=-(ML^mR_\a)
_-\hw=-\{\hw\sum_{k=1}^\infty kt_{k\a}(A\p)^{k+m-1}E_\a\hw\}_-\hw.$$
Here we concentrate on the case $m\leq 0$. The right hand side of the last
equation is a sum of three terms, with negative, zero and positive $k+m-1$, the
first of them is
$${\rm N}=-\hw\sum_1^{-m}kt_{k\a}(A\p)^{k+m-1}E_\a=-[\hw,t_{1\a}](A\p)^mE_
\a-\sum_1^{-m}kt_{k\a}\hw(A\p)^{k+m-1}E_\a.$$ (This term is absent for $m=0$).
The second is $${\rm Z}=-\{\hw(-m+1)t_{-m+1,\a}E_\a\hw^{-1}\}_-\hw$$
$$=-\hw(-m+1)t_{-m+1,\a}E_\a+(-m+1)t_{-m+1,\a}E_\a\hw$$ $$=-(-m+1)[\hw,t_{-m+1,
\a}E_\a]=-(-m+1)[\hw,t_{-m+1,\a}]E_\a\delta_{m,0}-(-m+1)t_{-m+1,\a}[\hw,E_\a].
$$ The third: $${\em P}=\sum_{-m+2}^\infty kt_{k\a}\p_{k+m-1,\a}\hw$$
$$=-[\hw,t_{1\a}]E_\a\d_{m,0}+(-m+1)t_{-m+1,\a}\p_{0\a}\hw.$$
Now, notice:
$$i)~\hw(A\p)\exp(A^{-1}xz)=\hw(z)\exp(A^{-1}xz),~~ii)~[A\p,t_{1\a}]E_\a=E_\a,$$
$$iii)~\hw(A\p)t_{1\a}\exp(A^{-1}xz)E_\a=\{t_{1\a}\hw(z)+\p_z\hw(z)\}\exp(A^
{-1}xz)E_\a,$$ or $$[\hw(A\p),t_{1\a}]\exp(A^{-1}xz)E_\a=\pz\hw\exp(A^{-1}xz)E_
\a.$$ Applying the operator $\p_{1,m,\a}^*\hw$ to $\exp(A^{-1}xz)$ and
canceling $\exp(A^{-1}xz)$ we get\\

{\bf Proposition 1.} {\sl The action of additional symmetries $\p_{1,m,\a}^*$
with $m\leq 0$ on $\hw$ is given by}
$$ \p_{1,m,\a}^*\hw(z)=-z^m\pz\hw(z)E_\a-\sum_1^{-m}kt_{k\a}z^{k+m-1}\hw(z)E_\a
$$ $$+\sum_{-m+1}^\infty kt_{k\a}\p_{k+m-1,\a}\hw(z)$${\sl or, in coordinates,}
$$ \p_{1,m,\a}^*\hw_{\g\b}(z)=-z^m\pz\hw_{\g\b}(z)\d_{\a\b}-\sum_1^{-m}kt_{k
\a}z^{k+m-1}\hw_{\g\b}(z)\d_{\a\b}$$ $$+\sum_{-m+1}^\infty
kt_{k\a}\p_{k+m-1,\a}
\hw_{\g\b}(z)\eqno{(9.3)}$$

{\bf Remark.} {\sl Like the action of the KP operators on $\hw$ (see remark 2
at the end of Sect. 2), the action of additional symmetries is defined up to
terms $\hw b$ where $b=\sum_1^\infty b_i\p^{-i}$ is a series with constant
diagonal matrices $b_i$.} \\

One can also obtain an action of $\p_{1,m,\a}^*$
on the Baker function $w=\hw\exp\xi(t,z)$. It is easy to see that
$$ \p_{1,m,\a}^*w=-z^m\pz wE_\a+\sum_{-m+1}^\infty kt_
{k\a}\p_{k+m-1,\a}w.$$ If $w$ is $w_V$ i.e. the Baker function determined by an
element $V$ of the Grassmannian, then $w\in V$, the last terms of the above
equality also belongs to $V$, and $\p_{1,m,\a}^*$ acts on $w$ as the operator
$-z^m\pz{\bf E_\a}$ modulo $V$, where ${\bf E_\a}w$, by definition, is $wE_\a$.

Values of $w$ for all $t$ span the subspace $V$. Action of $\p_{1,m,\a}^*$ on
an element of the Grassmannian $V$ is the action of this operator on elements
of $V$ modulo $V$, i.e. the mapping $V\rarr H/V$ (see e.g. [1]). Therefore, we
have independently proved the proposition of Sect. 8: Additional symmetries
$\p_{1,m,\a}^*$ act on elements of the Grassmannian as operators $-z^m\pz{\bf
E}_\a$.\\

In [1] it was shown how, using the relations of the type (9.1) and (9.2), to
transfer the action of $\p^*$s from $\hw$ to $\tau$. It was
done in both ways: the equivalence between some actions of $\p^*$ on $\hw$ and
on $\tau$ was proven. Now, for brevity, we do this in one way: show that some
given action of $\p^*$ on $\tau_{\a\b}$ implies (9.3).

{\bf Proposition 2.} {\sl The action of additional symmetries $\p_{1,m,\a}^*$
with $m\leq 0$ on $(\tau_{\g\b})$ is given by
$$\p_{1,m,\a}^*\tau_{\g\b}=\big(\sum_{k=-m+1}^\infty kt_{k\a}\p_{k+m-1,
\a}+\frac 12\sum_{k+l=-m+1}klt_{k\a}t_{l\a}\big)\tau_{\g\b}+c\tau_{\g\b}
\eqno{(9.4)}$$ where $t_{k\a}$ are arguments of this element $\tau_{\g\b}$
(Eqs (9.1)and (9.2) show that some of them can be shifted, and then we should
replace $t$ in this formula by a shifted argument)}.

{\em Proof.} Let $\tilde{\tau}_{\a\b}=\tau_{\a\b}(...,t_{s\g}-\d_{\g\b}\cdot
1/sz^s,...)$. Eq (9.4) implies that $$\p_{1,m,\a}^*\tilde{\tau}_{\g\a}=
[\sum_{k=-m+1}^\infty k(t_{k\a}-\frac1{kz^k})\p_{k+m-1,\a}+\frac12\sum_{k+l=
-m+1}kl(t_{k\a}-\frac1{kz^k})(t_{l\a}-\frac1{lz^l})]\tilde{\tau}_{\g\a}+
c\tilde{\tau}_{\g\a}$$
We have $$ \p_{1,m,\a}^*\hw_{\g\b}=\tau^{-1}\p_{1,m,\a}^*\tilde{\tau}_{\g\b}
-\tilde{\tau}_{\g\b}\tau^{-2}\p_{1,m,\a}^*\tau.$$ First we consider $\b\neq\a$.
$$\p_{1,m,\a}^*\tilde{\tau}_{\g\b}=(\sum_{k=-m+1}^\infty kt_{k\a}\p_{k+m-1,
\a}+\frac12\sum_{k+l=-m+1}klt_{k\a}t_{l\a})\tilde{\tau}_{\g\b}+c\tilde{\tau}_
{\g\b}$$ and, as it is easy to see,
$\p_{1,m,\a}^*\hw_{\g\b}=\sum_{k=-m+1}^\infty
kt_{k\a}\p_{k+m-1,\a}\hw_{\g\b}$ that coincides with the required by (9.3).

Now, let $\b=\a$. Then $$\p_{1,m,\a}^*\hw_{\g\a}=\tau^{-1}[\sum_{k=-m+1}^\infty
k(t_{k\a}-\frac1{kz^k})\p_{k+m-1,\a}+\frac12\sum_{k+l=-m+1}kl(t_{k\a}-\frac1{kz
^k})(t_{l\a}-\frac1{lz^l})]\tilde{\tau}_{\g\a}$$
$$-\tilde{\tau}_{\g\a}\tau^{-2}
\big(\sum_{k=-m+1}^\infty kt_{k\a}\p_{k+m-1,\a}+\frac 12\sum_{k+l=-m+1}klt_{k\a
}t_{l\a}\big)\tau$$ $$=\sum_{k=-m+1}^\infty kt_{k\a}\p_{k+m-1,\a}\tilde{\tau}_
{\g\a}-\sum_{k+l=-m+1}t_{k\a}z^{-l}\hw_{\g\a}-\frac m2z^{m-1}\hw_{\g\a}.$$ If
one takes into account that $$\tau^{-1}\sum_{k=-m+1}^\infty z^{-k}\p_{k+m-1,\a}
\tilde{\tau}_{\g\a}=z^m\pz\tilde{\tau}_{\g\a}/\tau=z^m\pz\hw_{\g\a}$$ and that
the term $(m/2)z^{m-1}\hw_{\g\a}$ is irrelevant according to the above remark,
this coincides with (9.3). This proves our proposition.  \\

Thus, the string equation is equivalent to the requirement that
$$\sum_{\a=1}^n\big(\sum_{k=-m+1}^\infty kt_{k\a}\p_{k+m-1,\a}+\frac
12\sum_{k+l
=-m+1}klt_{k\a}t_{l\a}\big)\tau_{\g\b}+c_{\a\b}\tau_{\g\b}=0~~\g,\b=1,...,n.$$
This is the so-called Virasoro constraint. It is clear that also higher
Virasoro constraints can be written in the same manner as in the scalar case
(see, e.g. [1]). We skip this.\\

\vspace{.2in}
{\bf Part 2. Modified KdV (mKdV).}\\

\vspace{.2in}
{\bf 10. Definition.} Let $v_1,...,v_n$ be independent generators of a
differential algebra ${\cal A}_v$. We shall also use $v_k$ with any integer
subscript $k$ assuming that $v_{k+n}=v_k$. Put
$$L_i:=(\partial +v_i)...(\partial +v_n)(\partial +v_1)...(\partial +v_{i-1})
=(\partial +v_i)...(\partial +v_{i+n-1})$$
and $B_k^{[i]}=(L_i^{k/n})_+,\:k\in{\bf Z},$ which is a $k$th order operator.
Each $L_i$ can also be represented as
$$ L_i=\partial^n+u_1^{[i]}\partial^{n-1}+...+u_n^{[i]}.$$
The coefficients $\{u_k^{[i]}\}$, $k=1,...,n$ are elements of ${\cal A}_v$
$$ \hspace{.2in} u_k^{[i]}=F_k(v_i,...,v_{i+n-1}),\:k=1,...,n.\eqno{(10.1)}$$
These functions define the Miura transformation. They determine an embedding
of the differential algebra ${\cal A}_{u^{[i]}}$ of all differential
polynomials in {$u_k^{[i]}$} into ${\cal A}_v$,
$${\cal A}_{u^{[i]}}\subset {\cal A}_v.$$
(If we wish $u_1^{[i]}=0$ then it must be $v_1+...+v_n=0$).

For each $L_i$ we can construct its KdV hierarchy
$$ \hspace{.2in}\partial_kL_i=[B_k^{[i]},L_i],\:k=1,2,...\:;\:\partial_k
=\partial /\partial t_k.\eqno{(10.2)}$$

{\bf Lemma.} $B_k^{[i]}(\partial +v_i)-(\partial +v_i)B_k^{[i+1]}$ {\sl (where,
by definition, $B_k^{[n+1]}=B_k^{[1]}$) are zero-order differential operators.
}\\

{\em Proof}. Starting from an obvious relation $L_i(\partial +v_i)=(\partial +v
_i)L_{i+1}$ we derive $L_i^{k/n}(\partial +v_i)=(\partial +v_i)L_{i+1}^{k/n}$
whence
$$ B_k^{[i]}(\partial +v_i)-(\partial +v_i)B_k^{[i+1]}=-(L_i^{k/n})_-(\partial
+v_i)+(\partial +v_i)(L_{i+1}^{k/n})_-.$$
The r.h.s. is a zero-order $\Psi$DO hence so is the l.h.s. On the other hand,
this is a differential operator. $\Box$    \\

{\bf Corollary.} {\sl The system of equation}
$$\hspace{.2in}\partial_kv_i=B_k^{[i]}(\partial +v_i)-(\partial +v_i)
B_k^{[i+1]},\:i=1,...,n \eqno{(10.3)}$$ {\sl makes sense.}\\

This system (for every fixed $k$) is called a modified KdV equation (mKdV).\\

{\bf Proposition.} {\sl The equation (10.3) is an extention to ${\cal A}_v$ of
every equation (10.2) given on ${\cal A}_{u^{[i]}}$ (i.e. for each fixed
$i=1,...,
n$)}.\\

{\bf Remark.} {\sl Differential algebras ${\cal A}_{u^{[i]}}$ do not coincide
i.e. elements of one of them are not, generally, differential polynomials of
another. A transition of one of them to another via ${\cal A}_v$ (i.e. by
integrating the Miura formulas with respect to $v_k$) is the B\"{a}cklund
transformation discovered by Adler [8].}\\

{\em Proof of the proposition.} Let $\partial_k$ be given on the generators
$v_k$ of ${\cal A}_v$ by (10.3). Then
$$\partial_kL_i=\partial_k(\partial +v_i)...(\partial +v_{i+n-1})=\sum_{l=1}
^{i+n-1}(\partial +v_i)...(\partial +v_{l-1})(B_k^{[l]}(\partial +v_l)$$ $$-
(\partial +v_l)B_k^{[l+1]})(\partial +v_{l+1})...(\partial +v_{i+n-1})=
B_k^{[i]}L_i-L_iB_k^{[i]}=[B_k^{[i]},L_i].$$

All the equations (10.2), for any $i$, are the same KdV system but they differ
by their embedding into ${\cal A}_v$. If a solution {$u_k$} to the KdV equation
is given we can take it for $u_k^{[i]}$ and solve Eq.(10.1) with respect to
{$v_k$}. Solutions depend on some arbitrary constants which may depend on the
time variable. The previous proposition means that the constants can be chosen
properly to make {$v_k$} satisfy the mKdV equation. (It suffices to find
$v_k$ in an initial moment $t=t_0$ and then take them as initial conditions
for the mKdV system. If we return from $v_k(t)$ to $u_k(t)$ solving Eq.(10.1)
in
the opposite direction, this solution to KdV must coincide with the original
since they coincide at the initial moment).

Thus $n$ solutions of the KdV equation correspond to any solution of mKdV. They
are, of course, not independent, but closely connected by the equation
$$ \hspace{.2in} L_i(\partial +v_i)=(\partial +v_i)L_{i+1}~~(L_{n+1}=L_1).
\eqno{(10.4)}$$ We call them compatible. \\

{\bf 11. $\tau$-functions.} Let us have n compatible solutions to KdV equation,
$L_i$. Operators $L_i^{1/n}$ are solutions to KP hierarchy, and a
$\tau$-function corresponds to each of them, $\tau_i$.                     \\

{\bf Proposition.} {\sl The corresponding solution to mKdV can be expressed in
terms of $\tau_i$ as}
$$ \hspace{.2in}v_i=\partial\log (\tau_{i+1}/\tau_i)~~(\tau_{n+1}=\tau_1).
\eqno{(11.1)}$$

{\em Proof}. We represent all the $L_i$ in the form of dressing;
$$L_i=\hat{w}^{[i]}(\p)\p^n(\hat{w}^{[i]}(\p))^{-1},\:\:\hat{w}^{[i]}(\p)=\sum_
0^{\infty}w_k^{[i]}\p^{-k},\:w_0^{[i]}=1.$$
$\hat{w}^{[i]}$ are unique to within multiplication on the right by constant
series $\sum_0^{\infty}c_k^{[i]}\partial^{-k},$ $c_0=1$. Equations
$$\hat{w}^{[i]}\partial^n(\hat{w}^{[i]})^{-1}(\partial +v_i)=(\partial +v_i)
\hat{w}^{[i+1]}\partial^n(\hat{w}^{[i+1]})^{-1}$$ imply $[\partial^n,(\hat{w}
^{[i]})^{-1}(\partial +v_i)\hat{w}^{[i+1]}]=0$. Then $(\hat{w}^{[i]})^{-1}
(\partial +v_i)\hat{w}^{[i+1]}=\partial\cdot\sum_0^{\infty}c_k^{[i]}\partial
^{-k},$  $c_k^{[i]}=const,\:c_0^{[i]}=1$.

Using the freedom of multiplication of $\hat{w}^{[i]}$ on the right by constant
series we can obtain a simpler formula $(\hat{w}^{[i]})^{-1}(\partial +v_i)
\hat{w}^{[i+1]}=\partial$ and
$$\partial +v_i=\hat{w}^{[i]}\partial(\hat{w}^{[i+1]})^{-1}=(1+w_1^{[i]}
\partial^{-1}+...)\partial (1-w_1^{[i+1]}\partial^{-1}+...)$$
which implies $v_i=w_1^{[i]}-w_1^{[i+1]}$. On the other hand, the Baker
function can be expressed in terms of the $\tau$-function
$$\hat{w}^{[i]}(z)=\tau^{[i]}(...,t_k-1/kz^k,...)/\tau^{[i]}(t)$$
whence $w_1^{[i]}=-\partial_1\tau^{[i]}/\tau^{[i]}=-\partial\log\tau^{[i]}$.
Now the required equality is obvious. $\Box$  \\

{\em Now we are going to give a construction of a set of compatible
$\tau$-functions within the framework of Segal-Wilson's theory.}\\

The original idea was given by Wilson [9].

We know that solutions to the KdV hierarchy are related to elements of a
submanifold of the Grassmannian, $Gr^{(n)}\subset Gr$ defined by
the property $z^nV\subset V$ for $V\in Gr^{(n)}$.

Now, we have $n$ operators $L_i$ corresponding to one solution of mKdV.
Therefore, our main objects now will be $n$-tuples of elements of the
Grassmannian, $V=(V_1,...,V_n)$. Of course, they cannot be independent since
$L_i$ are not. We will require the following property:
$$ zV_i\subset V_{i+1},~~i=1,...,n;~~(V_{n+1}=V_1).\eqno{(11.2)}$$

{\bf Example.} Let $H$ be the space $L_2$ on the circle $|z|=1$, and
$$V_i=\{f(z)=\sum_{-N}^{\infty}f_kz^k\:|\:f(a_l)=\epsilon^{-i}\alpha_lf(
\epsilon a_l);\:l=1,...,N,\:\epsilon^n=1\}. $$ Functions $f$ are supposed to be
prolonged into the circle and $a_l$ are distinct points, $0<|a_l|<1$, while
$\a_l$ are arbitrary non-zero numbers. It is easy to see that all properties
are
satisfied.  \\

Eq.(6) implies that $z^nV_i\subset V_i$, thus, $V_i\in$Gr$^{(n)}$, and there
are
corresponding $\tau$-functions $\tau_i$ and Baker functions $w_{V_i}=w_{V,i}$.
\\

{\bf Proposition.} {\sl There are operators $\p+v_k$ such that}
$$(\p+v_k)w_{V,k
+1}=zw_{V,k}.$$

{\em Proof}. Let $g=\exp\xi(t,z)$ where $\xi(t,z)=\sum_1^\infty t_kz^k$. By the
definition of a Baker function
$$ w_{V,k+1}=(1+a_1z^{-1}+...)g\in V_{k+1},$$ then
\begin{eqnarray*}
\partial w_{V,k+1}&=&(z+a_1+...)g\in V_{k+1},\\
w_{V,k}&=&(1+b_1z^{-1}+...)g\in V_k,\\
zw_{V,k}&=&(z+b_1+...)g\in zV_k\subset V_{k+1}
\end{eqnarray*}
whence
$$ \partial w_{V,k+1}-zw_{V,k}-(a_1+b_1)w_{V,k+1}=O(z^{-1})g\in
V_{k+1},$$ therefore this quantity is zero (since the Baker function is
unique), and $v_k=-(a_1+b_1)$. $\Box$\\

{\bf Corollary.} {\sl If $w_{V,k}=\hat{w}_{V,k}g$, where $\hat{w}_{V,k}=\sum_0^
{\infty}w_{k,i}\partial^{-i}$, then  }
$$ (\partial +v_k)\cdot\hat{w}_{k+1,W}=\hat{w}_{k,W}\cdot\partial.$$

This is nothing but the compatibility condition for mKdV. It implies
$$(\partial +v_i)...(\partial +v_{i+n-1})\hat{w}_i=\hat{w}_i\partial^n$$
i.e. $L_i=\hat{w}_i\partial^n\hat{w}_i^{-1}.$ The formula (11.1) with $\tau_i
=\tau_{i,W},\tau_{i+1}=\tau_{i+1,W}$ provides us with a solution to mKdV.\\

{\bf 12. mKdV as a reduction of mcKdV.} The relation (10.2), (10.3), (10.4) and
others can be represented in a matrix form (see also [11]). Let
$$Q=\left( \begin{array}{cccccc}0&\p+v_1& & & &0\\ & &\p+v_2& & & \\ & & &
\ddots& & \\ \cdot&\cdot&\cdot&\cdot&\cdot&\cdot\\ & & & & &\p+v_{n-1}\\\p+v_n&
 & & & &0\end{array}\right).$$ Then $$Q^n=L={\rm diag}(L_1,...,L_n)\eqno{(12.1)
}$$ where $L_i$ were defined above. (This is, actually, the Miura
transformation). Let $B_k={\rm diag}(B_k^{[1]},...,B_k^{[n]}).$
Then Eq. (10.2) and (10.3) can be written as $$ \p_kL=[B_k,L],~~\p_kQ=[B_k,Q].
\eqno {(12.2)}$$ (these are KdV and mKdV, correspondingly).
The operator $Q$  looks like the operator (4.1) of the mcKdV hierarchy $h=1$.
There are two distinctions: firstly, we have matrices $u_0$ of a very special
form and, secondly, $A$ is no more diagonal but
$A_{ij}=\delta_{i+1,j}$ (recall periodicity: $i+n$ is identified with $i$).
There are various possibilities to deal with such $A$. For example,
changing the basis, we can reduce $A$ to a diagonal form $diag(1,\epsilon,.
.., \epsilon^{n-1})$ where $\epsilon$ is an $n$th root of
the unit. We can also preserve the matrix $A$ as it is, replacing
spectral projectors $E_\a$ by $$E_{\a}=(1/n)\sum_r\e^{-\a r}A^r.$$ However,
this project also has its disadvantage: the matrices $E_\a$ are
complex. The best of all, and we do this further, is to preserve the same $A$,
but, instead of the spectral projectors, just to use powers $A^{-\a}$ of $A$.

We have to restrict some mcKdV flow to the matrices $Q=(\p+V)A$ where
$V=$diag$(v_1,...,v_n)$. The submanifold of these matrices among all
matrices $\L=A\p+U$ will be denoted by {\bf Q}.

Let us consider dressing: $A\p+U=\hat{w}\p A\hat{w}^{-1}$ i.e. $(A\p+U)
\hat{w}=\hat{w}\p A$, $\hat{w}=1+O(\p^{-1})$. The dressing operator
$\hat{w}$ is not unique: it can be multiplied on the right by monic series
in $\p^{-1}$ of the type $\sum a_{jk}\p^{-j}A^k$, coefficients are scalar
constants.

Define basic mcKdV flows as $$\p_{m\a}\L=-[(\L^mR_\a)_-,\L]$$ where
$R_\a=\hw A^{-\a}\hw^{-1}$. {\it It is easy to see that} $\p_{11}=\p$.\\

{\bf Lemma.} {\sl If $\L=Q\in{\bf Q}$ then a diagonal matrix can be chosen as
$\hw$.}\\

{\em Proof}. Rewrite the dressing equations in coordinates:
$$ (\p+v_i)\hat{w}_{i+1,j}=\hat{w}_{i,j-1}\p. \eqno{(12.3)}$$ The equality
$L^n\hat{w}=\hat{w}\p^n$ yields $(\p+v_i)...(\p+v_{i+n-1})\hat{w}_{ij}=
\hat{w}_{ij}\p^n$. This
implies that $\hat{w}_{ij}\p^n\hat{w}_{ij}^{-1}$ does not depend on $j$,
$\hat{w}_{ij}\p^n\hat{w}_{ij}^{-1}=\hat{w}_{ij_1}\p^n\hat{w}_{ij_1}^{-1
}$ whence $\hat{w}_{ij_1}^{-1}\hat{w}_{ij}$ commutes with $\p^n$ and is a
constant (series), $c_{ijj_1}$. We have now $\hat{w}_{ij}=c_{ijj_1}\hw_{ij_1}$.
Evidently, $c_{i,j,j+1}c_{i,j+1,j+2}=c_{i,j,j+2}$ etc. Another relation between
these
coefficients we obtain replacing in (12.3) $\hat{w}_{i+1,j}$ by $c_{i+1,j,j+1}
\hat{w}_{i+1,j+1}$ and $\hat{w}_{i,j-1}$ by $c_{i,j-1,j}\hat{w}_{ij}$.
This gives $c_{i+1,j,j+1}=c_{i,j-1,j}.$ Two obtained relations together mean
that coefficients can be represented as $c_{i,j,j+r}=c_{i-j,r}$. Then $\hat{w
}_{ij}=c_{i-j,r}\hat{w}_{i,j+r}$ and, in particular, $\hat{w}_{ij}=c_{i-j,i
-j}\hat{w}_{ii}$ which means that $\hat{w}=\hat{\hat{w}}\sum c_{r,r}A^r
$ where $\hat{\hat{w}}$ is the diagonal part of $\hat{w}$. This proves
the lemma, since there is a freedom to multiply $\hw$ by $(\sum c_{r,r}A^r)^
{-1}$.\\

{\bf Proposition.} {\sl The mKdV flow $\p_k$ on ${\bf Q}$ can be extended to
the mcKdV flow $\p_{kk}$ on the manifold of all operators $\L=\p A+U$, and
variables $t_k$ can be identified with $t_{kk}$.}\\

{\em Proof.} Given an mKdV flow $\p_kQ=[L_+^{k/n},Q]$ where $L=Q^n$.
Let us ``undress" operators $\Lambda=\p A+U$: $\Lambda=\p A+U=\hat{w
}\p A\hat{w}^{-1}$. The operator $\hat{w}$ can be chosen so that being
restricted to operators $Q$ it becomes diagonal (above lemma). Take the
following mcKdV flow: $\p_{kk}\Lambda=[(\L^kR_k)_+,\L]$. Restriction of this
flow to operators $Q$ is
$\p_{kk}Q=[(Q^kR_k)_+,Q]$. We have $Q^kR=\hat{w}A^k\p^kA^{-k}\hat{w}^
{-1}=\hat{w}\p^k\hat{w}^{-1}$. We also have $L=Q^n=\hat{w}\p^n\hat{w}^{-1}$.
Then $L^{k/n}=\hat{w}\p^k\hat{w}^{-1}$ (we use here the fact that $\hat{w}$ is
a diagonal matrix since, by definition, $L^{k/n}$ is a diagonal operator
whose $n$th power is $L^n$). Thus, $Q^kR_k$ can be replaced by $L^{k/n}$ and
the
mcKdV equation becomes $\p_{kk}Q=[(L^{k/n})_+,Q]$ i.e. it coincides with the
given flow, if the variable $t_{kk}$ of the mcKdV hierarchy is identified with
the variable $t_k$ of the mKdV hierarchy. This proves the proposition.

Thus, the flows $\p_{kk}$ respect ${\bf Q}$ where they coincide with $\p_k$,
mKdV flows. The others, $\p_{k\a}$, with $k\neq\a$, are transverse to
${\bf Q}$.\\

{\bf Remark.} {\sl An element of the vector Grassmannian corresponding to
a solution of mcKdV that reduces to $Q$ is a direct sum of elements $V_i$ of
the scalar Grassmannian we were talking about in the last section.}\\

{\bf 13. mKdV, additional symmetries, and the string equation.} The situation
is the following. Additional symmetries act on all operators $L_i$ but it is
not clear whether they act in a compatible way (see above) and, thus, can be
transferred on mKdV (and we believe they cannot). However, if there is an
embedding of mKdV operators $Q$ into mcKdV operators $\Lambda=\p A+U$ one can
consider the additional symmetry flows on their whole manifold. They
commute with mcKdV flows, this is exactly why they are symmetries, but they do
not preserve the submanifold of operators $Q$. Nevertheless, if an operator $Q$
is independent of some additional time variable, then it satisfies the mcKdV
string equation, and this equation remains invariant under the mKdV flow
(which is, in fact a restricted mcKdV flow).

Let us see what is this string equation like. It was written in Sect. 5; now
we have $h=1$, $Q$ plays the role of $L$, $M$ is $\hat{w}\Gamma\hat{w}
^{-1}$ with $\Gamma=\sum_{l=1}^\infty\sum_{\b=1}^nlt_{l\b}(A\p)^{l-1}A^{-\b}$,
and $A$ is the matrix $\delta_{i,i+1}$.

Consider the additional symmetry $$\p_{1,-n+1,0}^*Q=-[(MQ^{-n+1})_-,Q]. \eqno{(
13.1)}$$ We have $$[Q,M]=\hw[A\p,\sum_{l=1}^\infty\sum_\a lt_{l\a}(A\p)^{l-1}A^
{-\a}]\hw^{-1}=\hw[A\p_{11},\sum_{l=1}^\infty\sum_\a lt_{l\a}(A\p)^{l-1}A^
{-\a}]\hw^{-1}=1.$$ Therefore this symmetry can be rewritten as
$$\p_{1,-n+1,0}^*Q=Q^{-n+1}+[(MQ^{-n+1})_+,Q]=(1+[(MQ^{-n+1})_+Q^{n-1},Q])Q^{-n
+1}.$$ The right hand side of this equation does not belong to ${\bf Q}$ (The
operator contains negative powers of $\p$), thus,
this additional symmetry cannot be reduced to ${\bf Q}$. \\

{\bf Proposition 1.} {\sl The additional symmetry (13.1) induces additional
symmetries $\p_{1,-n+1}^*$ on the KdV operators $L_i$ (see [1])}.\\

{\em Proof}. The last equation implies
$$\p_{1,-n+1,0}^*Q^n=n+[(MQ^{-n+1})_+,Q^n].$$ Now, $$MQ^{-n+1}=\hw\sum_{l,\a}
lt_{l\a}(A\p)^{l-1}A^{-\a}(A\p)^{-n+1}\hw^{-1}.$$ Let us freeze all the
variables: $t_{l\a}=0$ except for those that preserve ${\bf Q}$ i.e. with
$l=\a$. Then on ${\bf Q}$ we have $$MQ^{-n+1}=\hw\sum_llt_{ll}\p^{l-1}\p^{-n+1}
\hw^{-1}$$ which coincides with $ML^{-n+1/n}$ where $M=\hw\sum_llt_l\p^{l-1}\hw
^{-1}$, $\hw$ being a diagonal operator. The resulting equation coincides with
$\p_{1,-n+1}^*L=[(ML^{-n+1/n})_+,L]+n$ which is the direct sum of symmetries
$\p_{1,-n+1}^*$ acting on the operators $L_1,...,L_n$. Exactly this was stated.
$\Box$\\

{\bf Definition.} {\sl The mKdV string equation is $\p_{1,-n+1,0}^*Q=0$ or} $$[
(MQ^{-n+1})_+Q^{n-1},Q]=-1.\eqno{(13.2)}$$ As a direct corollary of the above
proposition we have \\

{\bf Proposition 2.} {\sl The mKdV string equation (13.2) induces the
KdV string equation $$[(ML^{-n+1/n})_+,L]=-n$$ or, better to say, a string
equation for each of operators $L_i$.}\\

Another useful formula: $$\p_{1,-n+1,0}^*Q=Q^{-n+1}+\sum_{l\geq n}\sum_\a l
t_{l\a}\p_{l-n,\a}Q$$ or, preserving only the variables $t_{ll}$,
$$\p_{1,-n+1,0}^*Q=Q^{-n+1}+\sum_{l>n}lt_{ll}\p_{l-n,l-n}Q.$$ For the KdV
operators this is $$\p_{1,-n+1}^*L=n+\sum_{l>n}lt_l\p_{l-n}L.$$

{\bf References.}\\

1. Dickey, L. A., Additional symmetries of KP, Grassmannian, and the string
equation, Preprint, 1992, hep-th 9204092.

2. Anagnostopoulos, K. N., Bowick, M. J. and Schwarz, A., The solution space of
the unitary matrix model string equation and the Sato Grassmannian, preprint,
1991.

3. Dickey, L. A., On Segal-Wilson's definition of the $\tau$-function and
hierarchies AKNS-D and mcKP, to be published in Proceedings of the Colloquium
on Integrable Systems, Marseille, 1991.

4. Orlov, A. Yu., Shulman, E. I., Additional symmetries for integrable
equations and conformal algebra representations, Lett. Math. Phys., 12, 1986,
171.

5. Chen, H. H., Lee, Y. C. and Lin, J. E., On a new hierarchy of symmetries for
the Kadomtsev-Petviashvili equation, Physica D, 9D, n03, 1983, 439.

6. Dickey, L. A., Soliton equations and Hamiltonian systems, Advanced Series in
Mathematical Physics, vol. 12, World Scientific, 1991.

7. Mulase, M., Category of vector bundles on algebraic curves and infinite
dimensional Grassmannians, Internat. Journal of Math., 1, no3, 1990, 293.

8. Adler, M., On the B\"acklund transformation for the Gelfand-Dikii equations,
Comm. Math. Physics, 80, no4,1981, 517.

9. Wilson, G., Infinite-dimensional Lie groups and algebraic geometry in
soliton theory, Phil. Trans. Roy. Soc., London, A315, 1985, 393-404.

10. Hollowood, T., Miramontes, L., Pasquinucci, A. and Nappi, C., Hermitian
versus anti-Hermitian 1-matrix models and their hierarchies, Nucl. Phys.,
B373, 247, 1992.

11. Kupershmidt, B. A. and Wilson, G., Modifying Lax equations and the second
Hamiltonian structure, Invent. Math., 62, 403, 1981.

\end{document}